# A LOGICAL AND TOPOLOGICAL PROOF OF THE IRREDUCIBILITY OF CONSCIOUSNESS TO PHYSICAL DATA

Iegor Reznikoff
Professor Emeritus, Département de Philosophie,
Université de Paris Ouest, Nanterre, France
E-mail: dominiqueleconte@yahoo.fr

We show here that what we call 'visual space of consciousness', the space of what we see, is a specific space different from the purely physical one and that its properties imply that it cannot be reduced to or deduced from physical laws. Some biological points are also briefly considered. The arguments are of logical, mathematical and physical character, and although elementary they require a careful reading (A first shorter version of this paper appeared in a hardly accessible Journal [1], and presented at the International Congress of Logic, Methodology and Philosophy of Sciences, in Beijing, August 2007). There is no need to define consciousness; we only observe some of its properties, namely geometric and topological properties of visual consciousness, and show that these properties cannot be based on physics only. Now, if a part of consciousness cannot be grounded on physics only, it is the same for consciousness as a whole and we speak of the irreducibility of consciousness to physical data. We do not consider philosophical questions or issues; in a simple physical and mathematical frame we give a logical proof of this irreducibility. Elements for a formal mathematical, logical proof are mentioned at the end of the paper.

## I. INTRODUCTION

The main purpose of this work is to give a proof of the non reducibility of consciousness to physical data. In order to treat the problem precisely and have clear definitions we essentially limit the question to the *visual space* i.e., to the space we see (when looking at something), so that we need not define *consciousness*. If a part – the visual one – of consciousness cannot be founded on physics only, it is the same for consciousness as a whole and we speak of the irreducibility of consciousness to physical data. In Sections II and III, notions of *visual space* and *irreducibility* are respectively defined; then we study two main properties of the visual space, namely its *continuity* (Section IV) and *unity* (Section VI). Since regarding unity the biological level is concerned, this point is briefly discussed (Section VII). The last Section gives elements for a formal mathematical and logical proof (which is out of the scope of this paper). Finally, we conclude with a short historical perspective.

## II. THE CONSCIOUSNESS SPACE

For a given observer, let A be the space of 'physical reality' as known by physics, the 'real' space of matter with what is included in it: moving atoms, particles and waves; and let B be the observer's brain regarded as a space, with its physiological and neuronal activity (of course $B \subset A$); then let C be the space of the observer's perceptive consciousness: what he sees, hears, touches, etc. considered as a space. There are of course further levels of consciousness, in particular a witness consciousness: the one that sees, hears, etc (not to speak of the thinking one). But here we consider only perceptive consciousness, what is seen (heard, etc), and, more precisely, the visual space in the case of vision (resp. the spaces of what one hears, touches, etc. for the other senses). For simplicity and because of its obvious geometric appearance, we confine our remarks mostly to the visual space, but the same points can be made about other spaces of perception (For a very interesting approach of consciousness of sound, see [2] and [3]). In what follows we will speak of A, B, and C also as being respectively the '*real*' or *physical space*, the brain space and the *consciousness space* with its visual sub-space.

Between a part of A − B (elements of A that are not in B) and B there is a map, say $f$, which to physical events in this part of A − B, through the perceptive channels, associates reactions in the brain space B. For instance, a photon flux received by the eyes creates an activity in the optical nerve and then in the brain. To this activity at the level B corresponds in general a representation in C; let $g$ be this correspondence between brain activity and its representation (image) in the consciousness space. There is, therefore, a correspondence from A to C defined by $h = g[f]$. We have the following diagram (Figure 1).

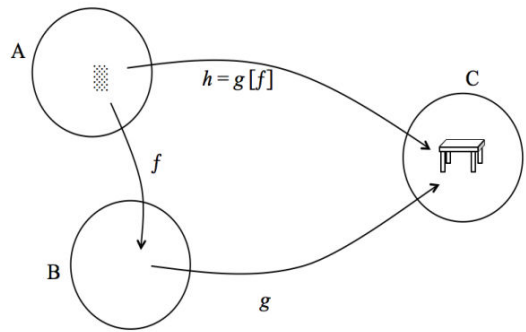

Figure 1. Spaces A, B, C as presented in the text.

If we limit C to perceptive consciousness – as we are doing here – we could expect to have g [$f$(A – B)] = C, but this is not the case since we are going to show that there are properties of C which do not proceed from properties of A. Note that $f$ is not injective, since different points in A cannot always be distinguished in B, and even less in C. One might discuss, of course, the precise domains of definition of $f$ and $g$ – since there are not, for all events in A or in B, corresponding reactions or representations in B or C, respectively, but this is of no importance here. Also we did not give a precise mathematical definition of the maps since it is not even clear what the elements in A, B and of course in C are; one should probably rather speak of a map between some subsets of A and subsets of C, or even better in the language of categories; for an approach of some properties of consciousness in terms of categories, see [4] (I am thankful to the mathematician G. Choquet who mentioned this remarkable work). For sake of simplicity, we keep the elementary formulation in the language of set theory, also because we shall have a logical approach referring to some axioms of set theory. The given correspondences $f$ and $g$ are however clear enough to say that $h$ gives a representation or an image of the reality A in the consciousness space C. For instance: to a subset T of A – B, a subset of particles reflecting a flux of photons, is associated, through the brain space B, a representation $h$(T) of T, in C, say a table; this table is a representation of external physical data (the set of all particles, waves, etc. concentrated in the given space we call *table*). We are going to study some properties of the space of such representations, what we call the consciousness space C.

*Proposition 1. The space C is a specific space in itself, different from A and B.*

Proof. We have C ≠ A, since the representation of A – B in C depends on B (and $g$), whereas A – B does not. For example if B is disturbed or injured, clearly so will be C (for instance a person sees two tables or none), while A – B has not changed (there is still one table). But also C ≠ B. What we see is not the brain, nor the activity of the brain. The same arguments show, moreover, that C ∩ A = ∅ and C ∩ B = ∅. Stated otherwise, the proposition says that
*the consciousness space is different from the real physical space and from the brain space.*

Given an object X in A, we do not see X in itself but only the result in C of a physiological and neuronal activity in B created by photons coming from X. This is well known, but the unconscious identification of $h$(X) with X, of what we see with the material physical world, is so strong and so widespread – even for those who have read Kant – that it is necessary to restate clearly: there is a proper space of consciousness and the picture that one sees is not the physical world. It is merely a representation of a set of particles and waves in the consciousness space; it does not mean, however, that this space is itself a 'set of particles': this is the point, the visual space is a proper non discrete space (see below).

An interesting question is that of the dimensionality of the consciousness space. For the visual space, one usually counts three spatial dimensions and three dimensions for colours. But we do not see a mixture of three colours; we see at least seven colours and their varieties and mixtures (For a topological approach of visual and colour spaces, see [5], [6], [7]). And what about the dimensions of the spaces of what we hear [3], smell or touch? Here, however we do not consider this question.

### III. ON IRREDUCIBLE PROPERTIES OF THE CONSCIOUSNESS SPACE

It is clear that the consciousness space has properties that are not as such in the physical space: this is the case, for instance, for colours, which are indefinable without the direct experience of seeing them (a person blind from birth can have no idea of what *green* means, though he might associate other feelings with this word). However, although the qualities (*qualias*) of a colour cannot be given physically, they do correspond to physical data: a green object reflects the light waves with a frequency that we see as 'green'. We say that this property of being green can be *reduced* to physical properties.

The question then arises whether there are properties of the consciousness space that cannot be reduced to real physical ones. If so, we speak of *irreducible* properties.

### IV. CONTINUITY OF THE CONSCIOUSNESS SPACE

A remarkable property of the visual space (but it is true also of other perceptions, although the matter is more difficult to formulate) is its continuity. We understand continuity in the mathematical sense, but the following elementary definition is sufficient here: in the visual space, there are no gaps or moving separate points; e.g., an ordinary white sheet of paper appears uniformly, permanently white and still, for at least a while. In contrast, physical reality at the atomic level is essentially discrete, non uniform, never motionless, and full of collisions; it doesn't mean that there are holes of energy or whatever, but we number atoms, electrons and various particles. With modern laser and other technologies, an isolated electron can be observed [8], [9], and with the NV (Nitrogen Vacancy) nanotechnology it is possible to produce sources of isolated photons [10]. Moreover, particles and waves are in perpetual movement. Of course, at our macroscopic level, we can use a magnifying glass and discover other aspects of what our eyes and consciousness did not see before, but the image remains continuous; there are no holes in the space of visual consciousness. Thus we have the question: how can a discrete moving atomic reality be represented in a continuous way (both in space and time)? The usual explanations about such questions concern the macroscopic level of perception. A typical example is given by the continuous appearance of a discrete pile of stones seen from a distance: it seems to be a white continuous spot. The usual argument is that the discreteness is too subtle to be perceived. The question then arises: to be perceived by what? And where? Indeed this gives no explanation at the atomic level, since the perceptions are transmitted and received (if we remain in a purely physical world) by discrete processes of particles and moving waves, particularly photons, and charges. How does this produce a continuous image, and where does this image appear? Certainly not in a physical space of particles, nor in the neuronal brain; the neurones transmit physical information up to consciousness which produces the continuous image we see; this is also a proof that C in no way belongs to the physical or brain space, and is a specific space of non-material, non-physical nature. Something to be perceived needs a perceiver and here the perceiver's visual continuous space cannot be reduced to physics because of the argument above. If everything were created, transmitted and received by physical spaces, it would remain permanently moving and discrete. As we say in French, 'the most beautiful girl in the world cannot give more than what she has': physics gives no more than physics.

*Proposition 2. The property of continuity of the consciousness space C is not reducible to or even explainable in terms of physical reality.*

Against this statement, there is also the argument that if there were holes in the visual space we simply would not see them, since we obviously can't see what we do not see (!). The discrete structure therefore cannot be perceived. However this argument again supposes that something is perceived and already presupposes a perceiver: who is the *we* in the sentence above and who or what is seeing? In such answers it is assumed that something (somebody) already sees or doesn't see; and the question remains how, in the final analysis, a 'continuous' space of vision can exist and where it can be located. Since it cannot be based on physics only, the conclusion is straightforward: continuity is a creation of consciousness. And here we come to a purely mathematical and logical consideration: continuity is not definable from discreteness and finite considerations, and cannot exist in a finite numbered domain. But physical reality – in a bounded domain at least – is finite. The property of continuity is, therefore, indeed irreducible to any physical reality, unless the notion of continuous field be introduced, which is a very theoretical and problematic notion that we discuss below.

*A. Commentary*

One can discuss whether the property of continuity is needed to characterize the visual space; for instance, isn't the property of *density* sufficient (as for the line of rational numbers)? Let us recall that density means that between two points there is always a third one. The above irreducibility argument remains in force even if we assume density; since density implies infinity, even in an interval or bounded space; indeed the fact that between two points there is a third one implies that between these two there is an infinity of points. An absolute proof of the continuity of the visual space in the strong mathematical meaning is certainly not possible because it requires high technical considerations of infinite character – let us recall that the continuity of a space implies that its infinity is not countable, which means that it is bigger than the infinity of the set of integers (it is said to have the power of the continuum). Such considerations are purely theoretical and certainly beyond any experience. But there is another strong epistemological argument for attributing continuity to the visual space. This argument comes from answering the question: how did the concepts of geometrical (Euclidian) space and precisely of continuity appear? How the geometrical line was and is understood to be continuous?

The notions of *geometrical space* and line appear of course in and from our visual space and visual experience (connected with that of movement and touch for three-dimensional awareness). Moreover, all our intuition of space geometry in the plane comes essentially from our visual space which, as we know, until the discoveries of Relativity Theory, was considered in its Euclidian formulation to be absolute (from the physical point of view at least). The notion of *continuity* proceeds as well from our visual experience, the best notion of a continuous line or surface being probably given by a surface of water: there are no holes or separations. That a segment of a straight line has infinitely many points (because it is dense) is readily understood and has been understood since Antiquity as well as the (intuitive) continuity of the line. And it is most remarkable that children, from visual experience, easily understand the notion of a (straight) line as well as its potential infinity and its continuity as being with 'points everywhere' so that there are no holes left. Of course the notions of closeness, or of going through are also related to our experience of movement and touch, but, finally it is by reasoning on the geometric line, which belongs to and comes from our visual space (so that it can be drawn), that the theory of this geometric line has been worked out. Also let us note that in our visual space all (necessary macroscopic) movements are continuous: it is impossible to join two points without passing somewhere in-between, while this is not the case at the atomic quanta level. It is therefore quite reasonable to consider that our intuition and understanding of the visual space *demonstrate it to be a continuous space*.

What we said about spatial continuity can be repeated concerning the continuity of the visual space in time (and, more generally, of perception in time). The visual space lasts in a continuous way as does a continuous movement; while at the atomic level, in duration, there is no continuity at all. However our perception of time is continuous and has led to a theoretical treatment of time which identifies it with the geometric line (for a study on time based on a distinction of physical and mind levels, see [11]). This continuity in time is closely related to what appears to be an even more remarkable property of consciousness and particularly of the visual space, namely its unity (see below, Section VI).

*B. The consciousness space as a field*

The only physical approach to continuity is given by the notion of field. For instance, an electromagnetic or gravitational field is assumed to be defined and active everywhere in the physical domain where it acts, and this *everywhere* is understood to be continuous since the space where the field is active is mathematically considered to be the three-dimensional space R3.

This is a purely mathematical and theoretical formulation: we can only verify that the field acts on every particle or object appearing in the domain, and experience can show no more. But to assume continuity and R3 allows us to use the mathematical infinitesimal calculus with all its tremendous power. However, we claim, after the discussion above, that this geometrical approach is a creation of consciousness and particularly of consciousness of the visual space, since there is no other evidence for such a geometrical and topological continuous conception. It is not and cannot be given by direct physical experience which is finite. For us, therefore, this geometry is not in A, but in C, and then induced from C to the theoretical, mathematical treatment by using the space R3 containing a theoretical model of A.

But since this notion of continuous field actually exists in physics – be it created by and conceived from visual consciousness – we may say that consciousness is indeed a field. And just as a movement of electrons creates an electromagnetic field, we may conjecture that intense brain activity – of purely physical character at the atomic level of particles and waves – creates a field of consciousness: the greater the brain activity, the richer the field of consciousness. This field is, of course, not physical, since $A \cap C = \emptyset$ as we have seen, the space C being a specific one. Nor it is simply reducible to a known physical field, we certainly do not see an electromagnetic or gravitational field. Moreover, in visual consciousness, we can isolate forms, colours, objects, etc., while even if there are different wave lengths, etc. in physical fields, it is consciousness that extracts the mentioned forms, properties or elements from the visual consciousness space we see. There is nothing analogous or even expressible concerning physical fields. The property of seeing separate objects in the unity of the whole visual picture corresponds to the *Comprehension Axiom* of set theory: given a set E and a property P, the subset of elements of E verifying the property P, exists. It cannot be deduced from physics. It seems to be a fundamental property of consciousness related to the *a priori* capacity of consciousness to pay attention. Moreover, if we consider the whole perceptive consciousness, there is no homogeneity between visual pictures, acoustic perception or touch.

If we consider the spaces of what we see, hear, smell, taste or touch, as different subspaces of the whole consciousness space, the non-homogeneity of these subspaces is a quite peculiar fact, bearing in mind that they are all produced by the same kind of neuronal activity, since there seems to be no difference between the neurones of different perception areas in the brain. How can the same kind of neuronal activity produce such different worlds of perception? This question could yield another proof of the irreducibility of the consciousness space to the physical one. Of course the scales of various physical data producing the perceptions are quite different e.g., the scales of light waves, molecules (for the smell) or sound waves, but this does not explain the complete non-homogeneity of the corresponding subspaces of consciousness, whereas they are held together in a remarkable unity: I smell the rose that I see. This is a specific property of consciousness.

But since the emergence of space consciousness comes mostly from an intense brain activity of quantum electromagnetic nature, the relationship between such a field and the field of consciousness has to be investigated not only for isolated phenomena (for instance the fact that different frequencies of light produce different colours in the visual consciousness) but in the whole. Why would not a special intense physical activity – in the brain – create a field of different – non-physical – nature? Clearly, the intermediate biological level appears as an essential one.

## V. THE OBJECTIVITY OF THE CONSCIOUSNESS SPACE C

A peculiarity of the consciousness space is that it can be studied essentially only from inside, by itself: only consciousness knows consciousness. And "if you want to know my consciousness, look in yours" sounds as a wise saying. Thus we come to this important statement:

*Proposition 3. The properties of C can be seen by everybody: its study is therefore perfectly objective.*

Note that here the word objective has the same meaning as in natural sciences, e.g., physics, since everything we know is known from our perceptive consciousness, and everything we look at – for instance the position of a needle in a measuring apparatus – is seen in our visual consciousness, that is in C. If two persons see the same object (the needle at a given position) it is because it is the 'same' object in their respective visual spaces. Although the meaning of the word *same* cannot be explained; this meaning is based on a universal understanding without which no communication would be possible. Here, appears the common but meaningless question whether we all really see the same colours or objects: is the red that I see the same as the one you see? The question is meaningless because it cannot be verified, but *the simpler the hypothesis, the better it is*, and the simplest is to consider that indeed we have essentially the same consciousness. But as mentioned above, philosophical discussions are not considered here.

## VI. UNITY OF THE VISUAL CONSCIOUSNESS SPACE

One of the most remarkable properties of consciousness space – and moreover difficult to understand – is its unity, that is the capacity that consciousness has to gather perceptions as a whole; from a multiplicity of independent nervous impulses and neuronal processes consciousness produces a unified whole. We do not have consciousness of separated elements, but always of a coherent whole, even when looking at an isolated object.

This unity principle is the following: given separate elements $x_1, ..., x_n$, it is the actual capacity to conceive their totality i.e. the set $\{x_1, ..., x_n\}$. It is remarkable that this corresponds to an axiom of set theory; logically this property is not reducible. It cannot be deduced simply from the existence of $x_1, \ldots, x_n$ as separate elements. Therefore, it is not physically explicable, unless of course it is implicitly assumed (which is often the case, for instance when one assumes that things are somehow and somewhere 'observed' before any consciousness has been introduced). In particular, the argument that unity results simply from the simultaneity of neuronal processes in some centre of the brain is doubly inconsistent. First because the notion of simultaneity is meaningless without the notion of *now* or the notion of *at the same time as*, which presupposes a reference and a clock and therefore an observer, i.e. a consciousness that grasps this simultaneity, this very notion introduces already an observer, it is not an absolute notion. And the second inconsistency is that simultaneity presupposes certainly the comparison of at least two elements and hence the notion of totality, be it only of the set $\{x_1, x_2\}$ as a whole. Therefore, to have the notion of simultaneity we already need that of unity; it is impossible to avoid circularity. The simultaneity of physical events is perhaps necessary for consciousness of unity but not sufficient to explain it.

But even if the notion of simultaneity is given, the probability that all the possible visible 'dots' of our visual neurology (e.g., retina) are grasped together in a coherent unity (their number can be estimated of the order of $10^7$), this probability is of order of $2^{(10^7)}$ (2 to the power of 10 to the power 7), which is well beyond any physical meaning even at the level of light-wave length. Unity cannot emerge 'by chance'; moreover, it is permanent, continuous in time. The probability for this continuing unity is physically without meaning.

This capacity of totalization, this gift of perceptive consciousness, is certainly one of its most important properties and unity may be the most characteristic property of consciousness. Consciousness unifies elements that otherwise are not related; from this comes what is called *meaning*. But we are not discussing this here any further.

As we have seen, the unity principle has, of course, no equivalent in physics; theoretically, it has to be borrowed from logic. The property of unity, say of visual consciousness and of the space C, is thus irreducible.

*Proposition 4. The unity of consciousness is not reducible to physical properties.*

The question then arises how far logical arguments can be used in physics, biology and matters of consciousness. But if one looks at a deductive science, rigorously founded, the logical and mathematical arguments are hitherto unavoidable. Of course a science can be very rich as a descriptive one, but the claim is now: is it possible to 'explain consciousness' by neuronal and finally from purely physical processes? Since in our attempt, *explain* means *deduce* or reduce, the argument needs to avoid circularity, therefore a careful logical examination is needed, which we have attempted above: the unity of the visual space cannot be reduced to or explained by physics without circularity since a notion of unity is needed beforehand.

## VII. BIOLOGICAL UNITY

If at some level the property of unity is needed and has to be introduced as such, then, it could be given already at a different level. It is natural to assume this unity, as we have seen, as one of the characteristic properties of consciousness, but it could be attributed already at the biological level. One often speaks of the 'unity of the cell'. Is it not at this elementary biological level that a principle of irreducible unity has to appear?

That such unity is necessary as a global *principle* in biology is simply shown by the same argument as the one given for the impossibility of a random unity of the visual space. Suppose a biological organism of about $10^9$ (of about 10 to the power of 9) components (e.g., molecules); the probability that all these components should behave *together* in the right way in order to constitute a biologically viable unity, this probability is at *least* of the order of $2^{(10^9)}$ (2 to the power of 10 to the power of 9), which is, as we have seen before, beyond any physical meaning even at the atomic level: the age of the universe would not be sufficient for even one cell to have a chance to exist, not to speak of a more complex organism.

However, even if a biological property should normally appear and be stated before properties of consciousness (and moreover could explain some of its aspects), we have a knowledge of the visual space, of its continuity and unity, certainly clearer, at least in its immediacy, than an as yet unformulated principle of unity in biology.

## VIII. FORMAL APPROACH

For a formal, strictly deductive logical approach, we need different levels of axioms, laws and data, so that the following levels have to be distinguished.

1. The Logical level needed for mathematics. This introduces axioms e.g., the *Axiom of Totality*: given $x_1, \ldots, x_n$ the set $\{x_1, \ldots, x_n\}$ exists. The *Comprehension Axiom*: given a set E and a property P, the subset of elements of E verifying the property P, exists. And finally an *Axiom of Infinity*.
2. The Mathematical level: theory of Real Numbers and Analysis. Logical axioms are intended for mathematical notions and reasoning.
3. The Physical level with its proper axioms and laws. The notions of continuity etc. are borrowed from the level 2.
4. The Biological level (this level is not really needed here).
5. The Consciousness level.

It is important to stress that we need not define consciousness (which would be a big challenge since consciousness is irreducible to other levels); we only observe some properties of visual consciousness, i.e. of what we see. But for continuity, unity and consciousness of seeing various objects, we need axioms analogous to the axioms above; these axioms are necessary to explain the mentioned properties of the visual space, and necessary for a deductive construction showing rigorously the irreducibility of consciousness to other levels. Since these axioms are not given by physics, clearly the level of consciousness is not at the physical level and cannot be deduced from physics.

## IX. Conclusion

That consciousness space is relatively independent from external physical reality is a classical statement. For Plato, Consciousness precedes Matter (as we learn from the *Timaeus*, 34c), the same for Indian classical religious philosophy; Kant's thoughts on this topic are well known (*Kritik der reinen Vernunft*), but it is worth quoting Berkeley: "*The proper objects of sight not without the mind; nor the images of anything without the mind*" and also "*Images in the eye are not pictures of external objects*" [12]. Here, we have simply shown that this relative independence of the consciousness space and its specific nature can be proved convincingly.